\documentclass[preprint,11pt]{elsarticle}
\usepackage[letterpaper, top=1.2in, bottom=1.2in, left=1.2in, right=1.2in]{geometry}

\usepackage{mathalfa}
\usepackage{tabularx}
\usepackage{amsmath,amssymb}
\usepackage[utf8]{inputenc}
\usepackage{amsmath}
\usepackage{amsthm}
\usepackage{amssymb}
\usepackage{extarrows}
\usepackage{mathtools}
\usepackage{adjustbox}
\usepackage{dcolumn}%
\usepackage{amsmath}
\usepackage[linesnumbered,ruled,vlined]{algorithm2e}
\usepackage{makecell}
\usepackage{amsfonts}
\usepackage{slashbox}
\usepackage{amssymb}
\usepackage{diagbox}

\usepackage{float}
\usepackage{verbatim}
\usepackage{bm}%
\newcommand{\bq}{\begin{equation}}
\newcommand{\eq}{\end{equation}}

\usepackage[hidelinks]{hyperref}
\usepackage{dcolumn}
\usepackage{bm}

\usepackage[table]{xcolor}

\newcommand{\RN}[1]{%
  \textup{\uppercase\expandafter{\romannumeral#1}}%
}

\begin{document}
\begin{frontmatter}

\title{A divergence-free constrained magnetic field interpolation method for scattered data}

\author[ORNL2]{M. Yang}
\author[ORNL2]{D. del-Castillo-Negrete \footnote{Corresponding author delcastillod@ornl.gov}}
\author[ORNL1]{G. Zhang}
\author[ORNL2]{M. T. Beidler}

\address[ORNL2]{Fusion Energy Division, Oak Ridge National Laboratory, Oak Ridge, TN.}
\address[ORNL1]{Computer Science and Mathematics Division, Oak Ridge National Laboratory, Oak Ridge, TN.}

\begin{abstract}
An interpolation method to evaluate magnetic fields given unstructured, scattered magnetic data is presented. The method is based on the reconstruction of the global magnetic field using  a superposition of orthogonal functions. The coefficients of the expansion are obtained by minimizing a cost function defined as the $L^2$ norm of the difference between the ground truth and the reconstructed magnetic field evaluated on the training data. The divergence-free condition is incorporated as a constrain in the cost function allowing the method to achieve arbitrarily  small errors in the magnetic field divergence. An exponential decay of the approximation error is observed and compared with the less favorable algebraic decay of local splines. Compared to local methods involving computationally expensive search algorithms,  the proposed method exhibits a significant reduction of the computational complexity of the field evaluation, while maintaining a small error in the divergence even in the presence of magnetic islands and stochasticity. Applications to the computation of Poincaré sections using data obtained from numerical solutions of the magnetohydrodynamic equations in toroidal geometry are presented and compared with local methods currently in use.
\end{abstract}


\tnotetext[fn1]{{\bf Notice}:  This manuscript has been authored by UT-Battelle, LLC, under contract DE-AC05-00OR22725 with the US Department of Energy (DOE). The US government retains and the publisher, by accepting the article for publication, acknowledges that the US government retains a nonexclusive, paid-up, irrevocable, worldwide license to publish or reproduce the published form of this manuscript, or allow others to do so, for US government purposes. DOE will provide public access to these results of federally sponsored research in accordance with the DOE Public Access Plan.}
\end{frontmatter}

\section{Introduction}

The accurate and efficient evaluation of magnetic fields is at the heart of many problems in plasma physics in general and controlled nuclear fusion in particular. For example, transport studies in magnetically confined fusion plasmas involve the integration of particles orbits that require the successive evaluation of the magnetic field along the trajectory. In most cases of practical interest, the magnetic field information is provided as a data set containing the components of the magnetic vector field at discrete spatial locations. In these cases, the magnetic field evaluation at arbitrary locations requires the use of numerical interpolation techniques that can significantly influence the transport calculations. For example,  the interpolation of electromagnetic fields when computing charge particles’ trajectories can deteriorate energy conservation \cite{Lalescu_etal_2013} and influence the effective accuracy of a solver \cite{lalescu2010implementation}. Of particular interest are interpolation errors in the divergence-free condition of the magnetic field, that can destroy the conservation of particle invariants \cite{mackay2006divergence} and the volume preservation of the magnetic field flow \cite{finn_chacon2005}. Also, as originally discussed in \cite{brackbill1980effect} small errors in the divergence-free condition can lead to large errors in the numerical solution of the magnetohydrodynamic equations. 

A natural approach to guarantee the divergence-free condition is to interpolate the magnetic potential and then reconstruct the magnetic field by applying the curl operator to the interpolated field. Among early works following this approach is Ref.~\cite{finn_chacon2005} that interpolates the vector potential using local tricubic splines and constructs the magnetic field by taking the analytic curl of the spline formulas, thus guaranteeing the divergence-free condition. The work in Ref.~\cite{Ravu_etal_2016} also follows the potential approach and presents improvements to the  computation of the inversion of the curl operator from magnetic grid-data,  and the reconstruction of the B field from the interpolated potential. In Ref.~\cite{mackay2006divergence}, given B on a discrete grid, the vector potential is constructed using Fourier transforms and then interpolated using cubic splines.

An alternative approach considers the use of divergence-free interpolation functions. For example, Ref.~\cite{mcnally2011divergence} discusses divergence-free matrix valued radial basis function interpolation. This approach can go beyond the preservation of the divergence-free condition and incorporate, for example, curl-free  interpolation functions of value in the case of current free magnetic fields. For the case of stellarator fields, this approach was followed in Ref.~\cite{probert2011high}  using expansions of fourth-degree vector polynomials (known as Maxwell elements)  that solves exactly the static current-free and the divergence-free Maxwell equations. In the method described in Ref.~\cite{SPIRAL_2013}, the interpolation in toroidal geometry is done by fitting the poloidal magnetic field  data at several toroidal planes using Chebyshev polynomials  and computing the toroidal field from the divergence free condition.  

Other interpolation methods, not constrained to the divergence-free condition, include the scheme presented in \cite{ pfefferle2014venus} based on Fourier reconstruction in the toroidal and poloidal directions and cubic spline in the radial direction of flux coordinate systems. The method proposed in Ref.~\cite{ garnett1992simple} assumes a local quadratic expansion of the magnetic field and uses least-squares to find the local expansion coefficients assuming a regular grid without imposing a divergence-free constrain. 

Although our main interest is in the interpolation of magnetic fields, the problem of interpolating divergence-free velocity fields in incompressible fluids is mathematically identical. As a result, significant efforts have also  been devoted to divergence-free methods in fluid mechanics. In Ref.~\cite{Yeung_1988} a study was presented of the role of interpolation on the Lagrangian statistics from numerically calculated velocity fields in incompressible fluid turbulence.  Reference \cite{ van2013optimal} discussed optimal interpolation schemes for particle tracking in fluid turbulence.    More recently,  Ref.~\cite{Tapley_etal_2022} studied the consequences of breaking the divergence-free condition in inertial  particle tracing studies in viscous flows.  

In this paper we present an alternative new method based on the construction of an approximate magnetic field using  a linear superposition of orthogonal functions given discrete magnetic field data. Once this is done, the interpolation reduces to the direct evaluation of the approximated field. The expansion coefficients are obtained by solving, in the least-squares sense, an over-determined linear system of equations. The solution minimizes an error cost function defined as the $L^2$ norm of the difference between the ground truth and the approximate magnetic field, as well as the divergence-free condition, on the training data set. The method is global in the sense that  uses all the data at once to minimize the overall error and the expansion functions are well-defined in the whole space. In this approach the divergence-free condition is incorporated as a ``soft" constrain in the cost function. Therefore, although the method can achieve arbitrary small errors in the magnetic field divergence, the divergence condition is not preserved exactly to machine precision.  Most importantly, contrary to previous methods that rely on structured data (e.g., given at the nodes of coordinate meshes) in the proposed method the magnetic data can be scattered, i.e., with no structure or order between their relative locations.

The interpolation problem of divergence-free fields is a complex problem and, depending on the application, computational resources and error tolerances, different methods offer specific advantages and face unique challenges.  Compared to methods not enforcing the divergence-free constrain, our method, as well as other divergence-free methods,  prevents the inadvertently  introduction of subtle critical numerical errors in particle tracking calculations. However, different to potential-based divergence-free methods, our approach directly interpolates the magnetic field bypassing the need to construct the vector potential (involving the nontrival inversion of the curl operator) and the need to compute derivatives (prone to numerical error) of the interpolated potential.   The basis functions used in our method can be easy to  compute well-known functions that best represent the geometry, symmetries, and boundary conditions of the problem.  This flexibility can offer computational advantages over methods requiring the numerical pre-computation of  divergence-free (e.g., “Maxwell elements”) functions. Also, in our method the computation of magnetic field derivatives (needed for example for guiding center trajectory computations) is exact and greatly simplified by the use of recursion relations and/or simple analytical formulas of the basis functions. Most importantly, the majority of the previously proposed methods require structured data sets. That is, training data given at the nodes of coordinate meshes. Going beyond this restriction, our method is mesh-free and can be used to interpolate unstructured (e.g., randomly scattered) training data sets. Once the expansion is obtained, the proposed method reduces  the interpolation problem to the evaluation of well-known global functions which is computationally more efficient than local interpolation methods involving computational expensive search algorithms to match desired evaluation points to local pre-computed interpolation coefficients. 

In the next section we formulate the divergence-free constrained interpolation problem and discuss the geometry and coordinate systems of interest. Section 3 introduces the global basis function expansion, formulates the least-squared problem including the divergence-free constrain, and discusses the algorithmic complexity of the proposed method. Section 4 presents applications of the method to synthetic data obtained by sampling an analytic model and data produced by the numerical solution of the magnetohydrodynamic equations in toroidal geometry of interest to magnetically confined fusion plasmas. Also, an error and computational complexity study is presented comparing the proposed method to local splines. Section 5 contains the conclusions. 

\section{Problem formulation}\label{sec:problem}
The proposed interpolation method can be implemented in any coordinate system. Here we consider toroidal coordinates commonly used in the study of magnetically confined fusion plasmas. In these coordinates,
the magnetic field has the general form 
\bq\label{mag_toro}
        \boldsymbol{\mathfrak{B}}(r, \theta, \zeta)=  \mathfrak{B}_r(r,\theta,\zeta) {\bm e}_r + \mathfrak{B}_\theta(r,\theta,\zeta) {\bm e}_\theta + \mathfrak{B}_\zeta(r,\theta,\zeta) {\bm  e}_\zeta \,,
\eq
with divergence 
\begin{equation}\label{free_toro}
\small
        \nabla \cdot \boldsymbol{\mathfrak{B}} =  \frac{1}{r(R_0+r\cos{\theta})}\left\{\frac{\partial}{\partial r}[r(R_0+r\cos{\theta})\mathfrak{B}_r]
        + \frac{\partial}{\partial \theta}[(R_0 + r\cos{\theta})\mathfrak{B}_{\theta}] + \frac{\partial}{\partial \zeta}(r\mathfrak{B}_{\zeta})\right\},
\end{equation}
where ${\bm e}_\zeta$, ${\bm e}_\theta$, ${\bm e}_r$ are the corresponding unit basis coordinate vectors.
The toroidal coordinates $(r,\theta, \zeta)$ are defined by $x=(R_0+r \cos \theta) \sin \zeta$, $y=(R_0+r \cos \theta) \cos \zeta$, and $z=r \sin \theta$, where $(x,y,z)$ are Cartesian coordinates,
$r \in [r_{\rm min},r_{\rm max}]$ is the minor radius, $\theta \in [0, 2 \pi)$ the poloidal angle, $\zeta=[0, 2\pi)$ the toroidal angle, and $R_0$ denotes the major radius. The cylindrical coordinates case is discussed in  \ref{sec:cylindrical}.

The goal is to find a function $\widehat{\boldsymbol{\mathfrak{B}}}$ that approximates the ground-truth magnetic field $\boldsymbol{\mathfrak{B}}(r, \theta, \zeta)$,
\begin{equation}\label{eq_rescale_toro}
\boldsymbol{\mathfrak{B}}(r, \theta, \zeta) \approx \widehat{\boldsymbol{\mathfrak{B}}}(r, \theta, \zeta) := \widehat{\mathfrak{B}}_r(r,\theta,\zeta) {\bm e}_r + \widehat{\mathfrak{B}}_\theta(r,\theta,\zeta) {\bm e}_\theta + \widehat{\mathfrak{B}}_\zeta(r,\theta,\zeta) {\bm  e}_\zeta \,,
\end{equation}
where $\widehat{\mathfrak{B}}_r$, $\widehat{\mathfrak{B}}_\theta$, $\widehat{\mathfrak{B}}_\zeta$ are approximations of ${\mathfrak{B}}_r$, ${\mathfrak{B}}_\theta$, ${\mathfrak{B}}_\zeta$ in Eq.~\eqref{mag_toro} subject to the constraint $\nabla \cdot \widehat{\boldsymbol{\mathfrak{B}}}=0$. 
Once this is accomplished, the interpolation problem is reduced to the direct evaluation of the constructed field $\widehat{\boldsymbol{\mathfrak{B}}}$.

The approximation will be built by learning from a training dataset with $N_{\rm train}$ samples, i.e., 
\bq\label{giventoro1}
\begin{aligned}
\boldsymbol{\mathcal{D}}_k^{\rm train} := \left\{(r_n, \theta_n, \zeta_n, \mathfrak{B}_k(r_n,\theta_n,\zeta_n))\;|\; (r_n, \theta_n, \zeta_n) \in \mathcal{S} \text{ for } n = 1, \ldots, N_{\rm train}\right\},\\
\end{aligned}
\eq
where the sub-index $k$ labels the radial, $k=r$,
poloidal, $k=\theta$, or toroidal, $k=\zeta$ component.
%
%
%
Note that the elements of $\mathcal{S}$ can be any set of $N_{\rm train}$ points scattered in the $(r,\theta,\zeta)$ space, 
not necessarily corresponding to the nodes of a coordinate grid. 


To simplify the representation of the divergence in Eq.~\eqref{free_toro} 
we introduce the rescaled magnetic field 
%
\bq\label{mag_toro2}
        {\bm B}(r, \theta, \zeta)= {B}_r(r,\theta,\zeta) {\bm e}_r + {B}_\theta(r,\theta,\zeta) {\bm e}_\theta +  {B}_\zeta(r,\theta,\zeta) {\bm  e}_\zeta\,,
\eq
\begin{equation}\label{mag_rescaled}
\begin{aligned}
        {B}_r(r, \theta, \zeta)& := (R_0+r\cos{\theta})\mathfrak{B}_r(r, \theta, \zeta),\\[4pt]
        {B}_\theta(r, \theta, \zeta)& := (R_0+r\cos{\theta})\mathfrak{B}_\theta(r, \theta, \zeta),\\[4pt]
        {B}_\zeta(r, \theta, \zeta) &:= \mathfrak{B}_\zeta(r, \theta, \zeta) \, ,
\end{aligned}
\end{equation}
for which the divergence-free condition reduces to
\bq\label{free_toro2}
\frac{\partial (r{B}_r)}{\partial r}
        + \frac{\partial {B}_\theta}{\partial \theta} + r\frac{\partial {B}_\zeta}{\partial \zeta} = 0 \, .
\eq
Taking advantage of this, our goal is to develop an approximation,
\bq\label{mag_toro3}
        \widehat{\bm B}(r, \theta, \zeta)= \widehat{B}_r(r,\theta,\zeta) {\bm e}_r + \widehat{B}_\theta(r,\theta,\zeta) {\bm e}_\theta +  \widehat{B}_\zeta(r,\theta,\zeta) {\bm  e}_\zeta\,,
\eq
for ${\bm B}$ (rather than approximating  $\boldsymbol{\mathfrak{B}}$ directly) subject to the constrain 
in Eq.~(\ref{free_toro2}). Once this is done, the approximation,
$\widehat{\boldsymbol{\mathfrak{B}}}$, of the original field can be obtained using the inverse of Eq.~\eqref{mag_rescaled},
\begin{equation}\label{approx1}
\begin{aligned}
    \widehat{\mathfrak{B}}_r(r, \theta, \zeta)& = \frac{1}{R_0+r\cos{\theta}}{\widehat {B}_r}(r, \theta, \zeta),\\[4pt]
    \widehat{\mathfrak{B}}_\theta(r, \theta, \zeta)& = \frac{1}{R_0+r\cos{\theta}}{\widehat {B}_\theta}(r, \theta, \zeta),\\[4pt]
    \widehat{\mathfrak{B}}_\zeta(r, \theta, \zeta)& =  {\widehat {B}_\zeta}(r, \theta, \zeta).
\end{aligned}
\end{equation}

\section{The divergence-free global least-squares method}\label{sec:ls}

The first step of the proposed divergence-free global least-squares (DivFree-GLS) method is to 
write 
%
\begin{equation}\label{eq_component}
\begin{aligned}
\widehat{B}_r (r,\theta,\zeta)& := \sum_{\mathbf{j}\,\in\, \mathcal{J}_r} \alpha^{r}_{\mathbf{j}}\; \psi_{\mathbf{j}}(r, \theta, \zeta), \\
\widehat{B}_\theta (r,\theta,\zeta)& := \sum_{\mathbf{j}\,\in\, \mathcal{J}_\theta} \alpha^{\theta}_{\mathbf{j}}\; \psi_{\mathbf{j}}(r, \theta, \zeta),\\
\widehat{B}_\zeta (r,\theta,\zeta)& := \sum_{\mathbf{j}\,\in\, \mathcal{J}_\zeta} \alpha^{\zeta}_{\mathbf{j}}\; \psi_{\mathbf{j}}(r, \theta, \zeta), 
\end{aligned}
\end{equation}
where $\mathbf{j} := (j_1, j_2, j_3)$ is a multi-index in the index sets $\mathcal{J}_r$, $\mathcal{J}_\theta$ or $\mathcal{J}_\zeta$, $\psi_{\mathbf{j}}(r, \theta, \zeta)$ are global basis function and $\alpha_{\mathbf{j}}^r$, $\alpha_{\mathbf{j}}^\theta$, $\alpha_{\mathbf{j}}^\zeta$ are the corresponding expansion coefficients associated with the index $\mathbf{j}$.


\subsection{The global basis expansion}\label{sec:basis}
The flexibility of the method allows the global basis function $\psi_{\mathbf{j}}(r, \theta, \zeta)$ to be chosen based on the geometry, boundary conditions, and symmetry of the problem. Here we use 
 orthogonal polynomials in $r$, and orthonormal trigonometric functions in $\theta$ and $\zeta$ which in addition 
exhibit excellent performance on QR decomposition and least-squares problems.
In particular, we write
\bq
\label{basis}
\psi_{\mathbf{j}}(r, \theta, \zeta) := P_{j_1}(r)F_{j_2}(\theta)F_{j_3}(\zeta),
\eq
where the functions $P_{j_1}(r)$ are orthogonal,  
\bq
\int_{-1}^{1}P_{j_1}(r)P_{j_1'}(r)dr = \frac{2}{2j_1 + 1} \delta_{j_1 j_1'}\, ,
\eq
Legendre polynomials with $j_1 \in \mathbb{Z}^+$ \cite{abramowitz1988handbook}.
Note that an affine transformation can be used to map the bounded domain $r \in [r_{\rm min},r_{\rm max}]$ into the $[-1,1]$ interval.
The functions $F_{j_2}$ and $F_{j_3}$  are orthonormal,
\bq
\frac{1}{2\pi}\int_{0}^{2\pi}F_{j_2}(\theta)F_{j_2'}(\theta)d\theta = \delta_{j_2 j_{2}'} \, \qquad 
\frac{1}{2\pi}\int_{0}^{2\pi}F_{j_3}(\zeta)F_{j_3'}(\zeta)d\zeta = \delta_{j_3 j_{3}'},
\eq
trigonometric functions, $F_{j_2}(\theta) = e^{i j_2 \theta}$, $F_{j_3}(\zeta) = e^{i j_3 \zeta}$. 

The multi-index sets $\mathcal{J}_k$, for $k=r$, $\theta$ and $\zeta$, 
can be specified using a full tensor product or a total degree method. 
Given the maximal order $J_1 \geq j_1$ of Legendre polynomials $P_{j_1}(r)$, and the maximal frequencies $J_2\geq j_2$ and $J_3\geq j_3$ of the trigonometric functions $F_{j_2}(\theta)$ and $F_{j_3}(\zeta)$,
the full tensor product method defines the 
multi-index set  as the set containing all possible combinations of indices
\begin{equation}\label{eq_ind}
\mathcal{J}^{\rm full-tensor}_k:= \{\mathbf{j} = (j_1,j_2,j_3) \,\,|\,\,  0\leq j_1\leq J_1, |j_2|\leq J_2, |j_3|\leq J_3\}.
\end{equation}
A potential drawback of this definition is that, if high order and/or high frequencies are needed, and if there are gaps in the spectrum, 
the size of $\mathcal{J}_k^{\rm full-tensor}$ can be unnecessarily large resulting on a significant increase of the 
computational cost of the algorithm. If this is the case, to strike a balance between accuracy and efficiency, the multi-index set can be defined using the total degree method
that includes only the combination of indices satisfying a weighted upper bound constrain
\begin{equation}\label{eq_total}
\mathcal{J}_k^{\rm total-degree}:= \{\mathbf{j} = (j_1,j_2,j_3) \,\,|\,\,  w_1 j_1 + w_2 |j_2| + w_3 |j_3| \leq J\},
\end{equation}
where the weights $w_1$, $w_2$, $w_3$ and the upper bound  $J$ are given parameters. 
Regardless of the method used, the number of elements of the set $\mathcal{J}_k$ will be denoted by $M_k$ and, according to Eqs.(\ref{mag_toro3}) and (\ref{eq_component}), the total number of terms used in the expansion of $\widehat{\bm B}$ is
$M_{\rm term} = M_{r} + M_{\theta} + M_{\zeta}$. The key aspect of the proposed DivFree-GLS method is that 
$M_{\rm term} \ll N_{\rm train}$ which, as discussed in further detail below, implies that the approximation $\widehat{\bm B}$ uses considerably fewer terms than local interpolation methods.


\subsection{The least-squares system with the divergence-free constraint}\label{sec:lsdiv}

The expansions coefficients are determined by imposing the equality of the ground-truth fields,
$B_r$, $B_\theta$, and $B_\zeta$ and the approximated fields $\widehat{B}_r$, $\widehat{B}_\theta$, and $\widehat{B}_\zeta$, at the training sets $\boldsymbol{\mathcal{D}}_r^{\rm train}$, $\boldsymbol{\mathcal{D}}_\theta^{\rm train}$, and $\boldsymbol{\mathcal{D}}_\zeta^{\rm train}$  subject to the divergence-free condition. 
%
Using Eq.~(\ref{eq_component}), for the radial component, $\widehat{B}_r$,  this implies
%
\begin{equation}\label{algo_21_a}
\begin{aligned}
\sum_{\mathbf{j}\,\in\, \mathcal{J}_r} \alpha^{r}_{\mathbf{j}}\; \psi_{\mathbf{j}}(r_n, \theta_n, \zeta_n) = B_r (r_n,\theta_n,\zeta_n), 
\end{aligned}
\end{equation}
$\text{ for }  n = 1, \ldots, N_{\rm train}$, 
which can be written in a matrix form 
\bq \label{mat_form}
{\bf A}_r{\bm \alpha}_{r} =  {\bm b}_r,
\eq
where ${\bf A}_r$ is a matrix of size $N_{\rm train}\times M_{r}$, 
${\bm b}_r$ is a column vector of size $N_{\rm train}$ and $\bm \alpha_r$ is a column vector of size $M_{{r}}$, i.e.,
\begin{equation}\label{e38}
\begin{aligned}
&{\bf A}_r := 
\begin{bmatrix}
\psi_{\mathbf{j}_1}(r_1, \theta_1, \zeta_1) &\cdots & \psi_{\mathbf{j}_{M_{r}}}(r_1, \theta_1, \zeta_1)\\
\psi_{\mathbf{j}_1}(r_2, \theta_2, \zeta_2) & \cdots & \psi_{\mathbf{j}_{M_{r}}}(r_2, \theta_2, \zeta_2)\\
 &  \ddots&    \\
\psi_{\mathbf{j}_1}(r_{N_{\rm train}}, \theta_{N_{\rm train}}, \zeta_{N_{\rm train}})& \cdots & \psi_{\mathbf{j}_{M_{r}}}(r_{N_{\rm train}}, \theta_{N_{\rm train}}, \zeta_{N_{\rm train}})\\
\end{bmatrix},
\\[10pt]
&{\bm b}_r := 
\begin{pmatrix}
{B}_r(r_1,\theta_1,\zeta_1) \\
{B}_r(r_2,\theta_2,\zeta_2) \\
\vdots\\
{B}_r(r_{N_{\rm train}}, \theta_{N_{\rm train}}, \zeta_{N_{\rm train}})
\end{pmatrix}, 
\qquad
{\bm \alpha}_{r} := 
\begin{pmatrix}
\alpha^{r}_{\mathbf{j}_1} \\
\alpha^{r}_{\mathbf{j}_2} \\
\vdots\\
\alpha^{r}_{\mathbf{j}_{M_r}}
\end{pmatrix}.
\end{aligned}
\end{equation}

Following the same procedure, we can get the corresponding 
linear systems for 
$\widehat{B}_\theta$, and $\widehat{B}_\zeta$, i.e.,
\begin{equation}\label{algo_21_b}
\begin{aligned}
\sum_{\mathbf{j}\,\in\, \mathcal{J}_\theta} \alpha^{\theta}_{\mathbf{j}}\; \psi_{\mathbf{j}}(r_n, \theta_n, \zeta_n) =  B_\theta (r_n,\theta_n,\zeta_n), 
\end{aligned}
\end{equation}
\begin{equation}\label{algo_21_c}
\begin{aligned}
\sum_{\mathbf{j}\,\in\, \mathcal{J}_\zeta} \alpha^{\zeta}_{\mathbf{j}}\; \psi_{\mathbf{j}}(r_n, \theta_n, \zeta_n) =  B_\zeta (r_n,\theta_n,\zeta_n), 
\end{aligned}
\end{equation}
and their matrix forms
\begin{equation}\label{algo_22}
{\bf A}_\theta {\bm \alpha}_{\theta} = 
{\bm b}_\theta,\qquad 
{\bf A}_\zeta {\bm \alpha}_\zeta = {\bm b}_\zeta,
\end{equation}
where ${\bf A}_\theta$, ${\bf A}_\zeta$, $\bm b_\theta$, $\bm b_\zeta$, $\bm \alpha_\theta$ and $\bm \alpha_\zeta$ can be defined in a similar way as in Eq.~\eqref{e38}.

%
The next step is to impose the divergence-free condition  in Eq.~\eqref{free_toro2} for the approximation 
$\widehat{\bm B}$, 
\begin{equation}\label{div_eq}
 \begin{aligned}
r\frac{\partial \widehat{B}_r}{\partial r}  + {\widehat{B}_r} + \frac{\partial \widehat{B}_\theta}{\partial \theta} + r\frac{\partial \widehat{B}_\zeta}{\partial \zeta} = 0.
\end{aligned}
\end{equation}  
Using Eqs.~(\ref{eq_component}), \eqref{basis},  and the properties of the derivatives of the Legendre polynomials and trigonometric functions,
\begin{equation}\label{div}
\begin{aligned}
 \left. r \frac{\partial}{\partial r}{\widehat{B}_r}+
{\widehat{B}_r} \right|_{(r_n,\theta_n,\zeta_n)} & = \sum_{\mathbf{j}\,\in\, \mathcal{J}_r} \alpha^{r}_{\mathbf{j}}\left\{\left[ r_n  \frac{dP_{j_1}(r_n)}{dr}+ P_{j_1}(r_n)\right ] F_{j_2}(\theta_n)F_{j_3}(\zeta_n) \right\},\\
\left. \frac{\partial}{\partial \theta}{\widehat{B}_\theta} \right|_{(r_n,\theta_n,\zeta_n)} & =\sum_{\mathbf{j}\,\in\, \mathcal{J}_\theta} \alpha^{\theta}_{\mathbf{j}} \left[i\, j_2\, P_{j_1}(r_n) F_{j_2}(\theta_n)F_{j_3}(\zeta_n) \right],\\
\left. r \frac{\partial}{\partial \zeta}{\widehat B_\zeta} \right|_{(r_n,\theta_n,\zeta_n)}& =\sum_{\mathbf{j}\,\in\, \mathcal{J}_\zeta} \alpha^{\zeta}_{\mathbf{j}} \left[  i\, j_3 \,r_n P_{j_1}(r_n)F_{j_2}(\theta_n)  F_{j_3}(\zeta_n) \right ] \, .
\end{aligned}
\end{equation}
%
Substituting Eq.~\eqref{div} in Eq.~\eqref{div_eq}, the divergence-free constrain can be written in the matrix form
\begin{equation}\label{div_mat}
 \begin{aligned}
 {\bf A}'_r  {\bm \alpha}_{r} + {\bf A}'_\theta {\bm \alpha}_\theta + {\bf A}'_\zeta {\bm \alpha}_\zeta = \bm 0 \, ,
\end{aligned}
\end{equation}   
where the elements of the matrices ${\bf A}'_r$, ${\bf A}'_\theta$ and ${\bf A}'_\zeta$  are constructed using Eq.~\eqref{div}. 

The problem is thus reduced to the solution of the system of linear equations in Eqs.~\eqref{mat_form},
\eqref{algo_22} and \eqref{div_mat}. However, in general this system
is over-determined and therefore we seek a solution in the least-squares sense
\cite{bjorck1996numerical}. That is, we define
\begin{equation}\label{mat_con1_2}
{\cal E}({\bm \alpha}_{r},{\bm \alpha}_{\theta},{\bm \alpha}_{\zeta}) :=\left\|
\begin{pmatrix}
{\bf A}_r & {\bf 0} & {\bf 0}\\
{\bf 0} & {\bf A}_\theta & {\bf 0}\\
{\bf 0}& {\bf 0} & {\bf A}_\zeta \\
{\bf A}'_r & {\bf A}'_\theta & {\bf A}'_\zeta 
\end{pmatrix} 
\begin{pmatrix}
{\bm \alpha}_r \\
{\bm \alpha}_{\theta} \\
{\bm \alpha}_{\zeta}
\end{pmatrix} -
\begin{pmatrix}
{\bm b}_r \\
{\bm b}_\theta \\
{\bm b}_\zeta \\
{\bf 0}
\end{pmatrix}\right \|\, ,
\end{equation}
where
$\|u\| := \sqrt{\sum_k u_k^*u_k}$, is the $L^2$ norm 
and find  ${\bm \alpha}_r$, ${\bm \alpha}_\theta$, and ${\bm \alpha}_\zeta$ by minimizing ${\cal E}$. In the calculations presented here the minimization is performed  using  the  Matlab build-in function ``mldivide'' \cite{higham2016matlab}.
However, due to the sparsity of the system matrix, which we use to 
reduce the memory requirements,
other  efficient iterative methods  could also be used \cite{reich2003iterative,fong2011lsmr,powell1972rapidly}.

Once the coefficients ${\bm \alpha}_r$, ${\bm \alpha}_\theta$, and ${\bm \alpha}_\zeta$ are found,
we use Eq.(\ref{eq_component}) to construct $\widehat B_r$, $\widehat B_\theta$ and $\widehat B_\zeta$ and the transformation in Eq.(\ref{approx1}) to get the interpolated magnetic field $\widehat{\boldsymbol{\mathfrak{B}}}$ as a series expansion of the basis functions $\psi_{\bf j}(r,\theta, \zeta)$. At this point, if needed, a further reduction on the number of terms in the expansion can be performed by neglecting those terms for which 
\bq
\label{thr_deff}
{\bf \alpha}^k_{\bf j} \leq \epsilon_{\rm th} \|\boldsymbol{\mathfrak{B}}\|_{<>} \, ,
\eq
where  $\epsilon_{\rm th} \ll 1$ is a threshold parameter and 
\begin{equation}
\label{norm_deff}
\|\boldsymbol{\mathfrak{B}}\|_{<>} := \sqrt{ \langle \mathfrak{B}^2_{r,n} \rangle +
\langle \mathfrak{B}^2_{\theta,n} \rangle + \langle \mathfrak{B}^2_{\zeta,n} \rangle} \, ,
\end{equation}
with $\mathfrak{B}_{k,n}=\mathfrak{B}_{k}(r_n,\theta_n,\zeta_n)$, where as before the index $k$ is used to label the $r$, $\theta$ or $\zeta$ components, and $\langle f \rangle = 1/N_{\rm train} \sum_{n=1}^{N_{\rm train}} f_n$ is the ensemble average over the training set.
\subsection{Algorithmic complexity}\label{sec:complex}


The computational complexity of an interpolation method consists of two parts, the ``offline complexity" and the ``online complexity". The offline complexity represents the floating point operations per second (FLOPS) for computing the coefficients of the interpolation. Given $N_{\rm train}$ data points, the local cubic spline interpolation has $4\times N_{\rm train}$ unknown coefficients. Since an $N_{\rm train} \times N_{\rm train}$ linear system needs to be solved, the corresponding complexity cost of spline interpolation is of the order $\mathcal{O}(N_{\rm train}^3)$ \cite{golub2013matrix}. Although advanced algorithms, e.g., \cite{coppersmith1987matrix}, have been proposed to overcome this unfavorable scaling, they have limited practical use.
On the other hand, in the proposed DivFree-GLS method 
the unknowns ${\bm \alpha}_r$, 
${\bm \alpha}_\theta$, and ${\bm \alpha}_\zeta$ are obtained from the linear system in Eq.\eqref{mat_con1_2} with an algorithmic complexity of at most $\mathcal{O}(N_{\rm train}\times M_{\rm term}^2)$, which is significantly less than the algorithmic complexity of the local splines because
$M_{\rm term} = M_{r} + M_{_\theta} + M_{\zeta}\ll N_{\rm train}$ where $M_k$ denotes the number of elements in the  multi-index set $\mathcal{J}_k$.


The online complexity is the computational cost for evaluating the approximated magnetic field. To evaluate $\widehat{\boldsymbol{\mathfrak{B}}}$ at a discrete set of random spatial locations $\{(r_n,\theta_n,\zeta_n) \,| \, n = 1, \ldots, N_{\rm test}\}$, the cubic spline interpolation method has to either search the corresponding basis for each point $(r_n,\theta_n,\zeta_n)$, $n = 1, \ldots, N_{\rm test}$ or use all training data as the interpolation basis, which implies a complexity cost of the order $\mathcal{O}(N_{\rm test}\times N_{\rm train})$. In contrast, for the DivFree-GLS method,the complexity cost of evaluating the reconstructed magnetic field at  $\{(r_n,\theta_n,\zeta_n) \,| \, n = 1, \ldots, N_{\rm test}\}$ is  $\mathcal{O}(N_{\rm test}\times M_{\rm term})$, which is independent of  $N_{\rm train}$ and much smaller than for the local splines method because 
$M_{\rm term} \ll N_{\rm  train}$. 

\section{Numerical examples}\label{sec:exam}
This section presents three numerical examples. The first example uses synthetic data generated using an analytical field to demonstrate the accuracy and efficiency of the method compared to the cubic spline interpolation method. The second and third examples consider magnetic field data obtained with the extended-magnetohydrodynamic codes and compare the Poincar{\'e} sections computed using the proposed DivFree-GLS method with currently used libraries. 
In all the examples presented, the DivFree-GLS method is tested using scattered data. Practically indistinguishable results are obtained when using structure data in the DivFree-GLS method. On the other hand,when using spline-based local methods to benchmark and compare our results, we use  structure data since these methods cannot incorporate scattered data.

\subsection{An analytic magnetic field model}\label{secex:analy}
We consider a magnetic field model in toroidal coordinates given by
\bq\label{ex1_mag}
\boldsymbol{\mathfrak{B}}=\mathfrak{B}_\zeta(r,\theta) {\bf e}_\zeta + \mathfrak{B}_\theta(r,\theta) {\bf e}_\theta + \delta \boldsymbol{\mathfrak{B}}(r,\theta,\zeta), 
\eq
with toroidal equilibrium field 
\bq
\mathfrak{B}_\zeta(r,\theta)=\frac{\mathfrak{B}_0}{1 + (r/R_0) \cos \theta},
\eq
and poloidal equilibrium field 
\bq
\label{poloidal}
\mathfrak{B}_\theta(r,\theta)=\frac{-r}{\left( R_0+ r \cos \theta\right )} \frac{\mathfrak{B}_0}{q(r)} \, , \qquad q(r)=q_0\left(1+\frac{r^2}{\varepsilon^2}\right) \, ,
\eq
where $q_0$, $\varepsilon$, $R_0$ and $\mathfrak{B}_0$ are constants. 
The magnetic field perturbation, $\delta \boldsymbol{\mathfrak{B}}$, is given by 
\bq
\delta \boldsymbol{\mathfrak{B}}=\sum_{mn} \nabla \times {\bm A}_{mn} \, , \qquad {\bm A}_{mn}=\epsilon_{mn} f(r) g(r) h(r)\, \cos(m \theta + n \zeta)\, {\bf e}_\zeta \,,
\eq
with
\bq
f(r)=\frac{1}{2}\left[ 1 - \tanh \left( \frac{r-a}{l_{mn}}\right) \right ]\, , \qquad 
g(r)= \left( \frac{r}{r_{mn}^*}\right)^m \, ,
\eq
\bq
\label{h_fcn}
h(r)=\exp\left[
-\frac{\left(r-r_{mn}\right)^2}{2 \sigma_{mn}^2}
+\frac{\left(r^*_{mn}-r_{mn}\right)^2}{2 \sigma_{mn}^2}
\right] \, ,
\eq
where $r_{mn}=r^*_{mn} - m \frac{\sigma^2_{mn}}{r^*_{mn}}$, 
$r^*_{mn} = \varepsilon \sqrt{\frac{m}{n q_0}-1}$, and $\epsilon_{mn}$, $\sigma_{mn}$ and $l_{mn}$ are constants.
For the parameters of the equilibrium model we will choose
$\mathfrak{B}_0=2$, $R_0=1.5$, $a=0.5$, $q_0=1$, and $\varepsilon=a/\sqrt{6}$. 
For the perturbation parameters we will consider two cases:
(i) A one-mode perturbation with $(m,n)=(2,1)$, $\epsilon_{21}= 1 \times 10^{-4}$, $l_{21}=0.05$, and $\sigma_{21} = 0.1$; and (ii) A two-mode perturbation with $(m,n)=\{(2,1), (3,1)\}$ $\epsilon_{21}=\epsilon_{31}=2 \times 10^{-4}$, 
$l_{21}=l_{31}=0.05$,  and $\sigma_{21}=\sigma_{31}=0.1$.
%
The training datasets of magnetic field values, $\boldsymbol{\mathcal{D}}_r^{\rm train}$, $\boldsymbol{\mathcal{D}}_\theta^{\rm train}$, $\boldsymbol{\mathcal{D}}_\zeta^{\rm train}$ in Eq.~(\ref{giventoro1}), are obtained by evaluating the analytical field at $N_{\rm train}$ spatial locations. 


\subsubsection{Numerical error and computational complexity}

Given a sample test set $\{(r_n,\theta_n,\zeta_n)\,|\, n = 1,\ldots, N_{\rm test}\}$, we define the normalized point-wise error, 
\bq\label{err_pnt_def}
\Delta \mathfrak{B}_{k,n}:=\frac{|\mathfrak{B}_{k,n}-\widehat{\mathfrak{B}}_{k,n}|}{\sqrt{\langle \mathfrak{B}_{k,n}^2 \rangle}} \, ,
\eq
and the global approximation error,
\begin{equation}
\label{err_glob_def}
{\rm Err} := \|\boldsymbol{\Delta  \mathfrak{B}}\|_{<>} := \sqrt{ \langle \Delta \mathfrak{B}^2_{r,n} \rangle +
\langle \Delta \mathfrak{B}^2_{\theta,n} \rangle + \langle \Delta \mathfrak{B}^2_{\zeta,n} \rangle} \, ,
\end{equation}
where
$\langle f \rangle = 1/N_{\rm test} \sum_{n=1}^{N_{\rm test}} f_n$ is the ensemble average over the sample test set, 
$\mathfrak{B}_{k,n}=\mathfrak{B}_{k}(r_n,\theta_n,\zeta_n)$ and
$\widehat{\mathfrak{B}}_{k,n}=\widehat{\mathfrak{B}}_{k}(r_n,\theta_n,\zeta_n)$, where as before the index $k$ is used to label the $r$, $\theta$ or $\zeta$ components. 

%

%
For the approximation error analysis in the DivFree-GLS method we use the total degree method in Eq.~\eqref{eq_total} with $\{w_1,w_2,w_3\} = \{1,3,3\}$,
and vary the number of terms in the expansions, $M_{\rm term}$, by setting the parameter $J = 3,6,9,12,15$,
keeping $N_{\rm train}=64^3$ fixed. In addition, we further reduced the terms in the expansion by using the threshold condition in  Eq.(\ref{thr_deff}) with $\epsilon_{\rm th}=10^{-8}$.

To illustrate the mesh-free properties of the   DivFree-GLS method, in the calculations shown, the $N_{\rm train}$ points were randomly scattered. Practically indistinguishable results (not shown) were obtained using data on an structured, coordinate grid. 
For the splines method we considered structured data and varied the number of $N_{\rm train}$ data values on a $\{(r_n,\theta_n,\zeta_n)$, $n=1, \dots N_{\rm train}\}$ grid. 

%

\begin{figure}[h!]
\center
    \includegraphics[scale =0.5]{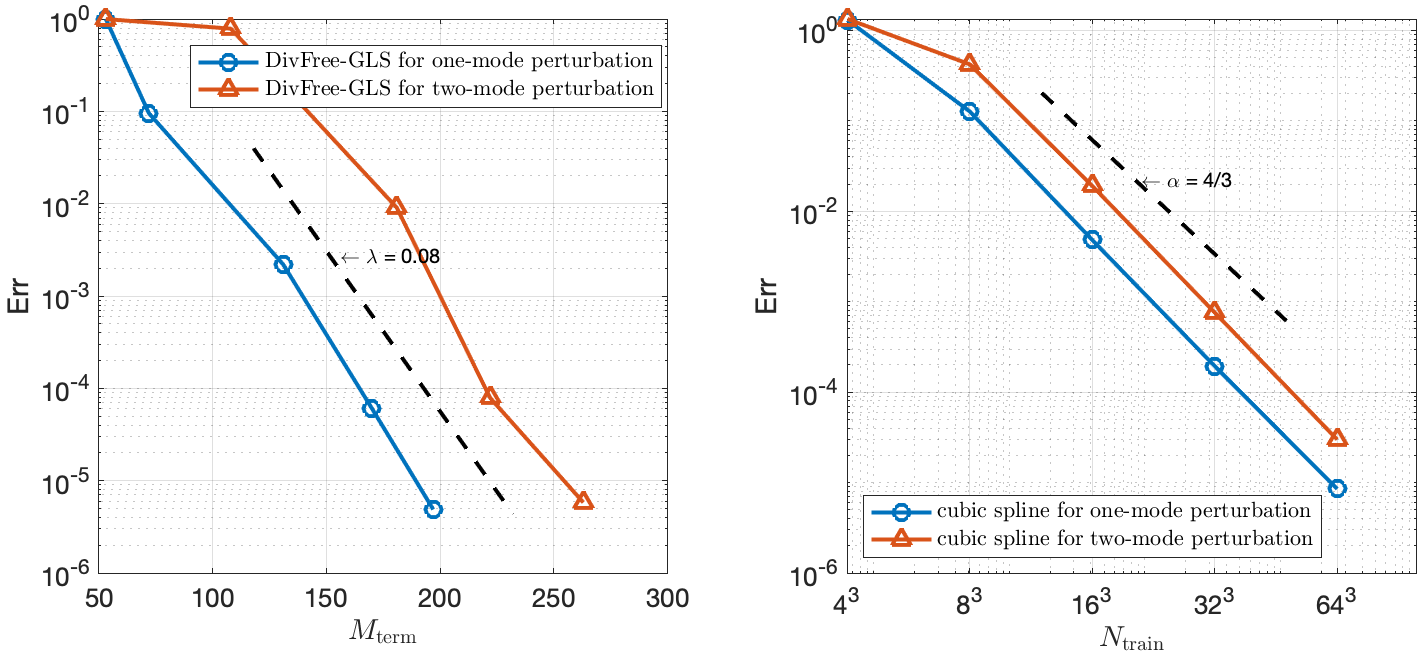}
    \vspace{-0.4cm}
    \caption{Global approximation error Eq.~(\ref{err_glob_def}) for the two cases of the analytic magnetic field example in Eq.~(\ref{ex1_mag}).  In the  DivFree-GLS method, shown in the left panel, the global error  decays 
    {\em exponentially} with $M_{\rm train}$ for mesh-free randomly scattered $N_{\rm train}=64^3$ data. Practically identical results (not shown) are obtained for data on an structured grid. The dashed line shows an exponential fit of the form ${\rm Err}\sim e^{-\lambda M_{\rm term}}$ with  $\lambda \approx 0.08$. 
    On the other hand, the local splines method, shown on the right panel, exhibits a less favorable {\em algebraic} decay with  $N_{\rm train}$. The dashed line shows an algebraic fit of the form ${\rm Err}\sim N_{\rm train}^{-\alpha}$ with $\alpha \approx 4/3$. Since local splines are limited to structured data, the training data sets correspond to nodes 
    on a $\{(r_n,\theta_n,\zeta_n)$, $n=1, \dots N_{\rm train}\}$ coordinate grid. 
}
    \label{error_ex1}
\end{figure}

Figure \ref{error_ex1} compares the decay of the global approximation error ${\rm Err}$ in Eq.(\ref{err_glob_def}) in the DivFree-GLS and the cubic spline interpolation methods. 
The DivFree-GLS method, shown in the left panel, exhibits an {\em exponential}-type decay of ${\rm Err}$ with the number of terms in the expansion, $M_{\rm term}$. On the other hand, the cubic spline interpolation method, shown in the right panel, exhibits a less favorable {\em algebraic}-type decay with the number of training points, $N_{\rm train}$. The key point is that to increase the accuracy of the spline method, $N_{\rm train}$ needs to increase considerably. But,  
the accuracy of the DivFree-GLS method can be improved by modestly increasing $M_{\rm term}$, keeping $N_{\rm train}$ fixed.


\begin{figure}[h!]
\center
    \includegraphics[scale =0.5]{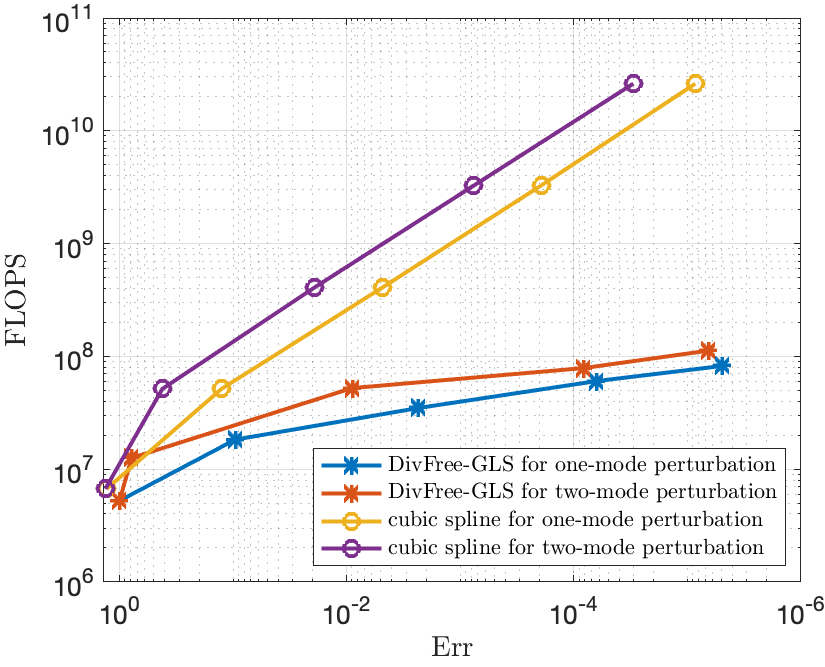}
    \vspace{-0.4cm}
    \caption{Online computational cost of the DivFree-GLS and the cubic spline interpolation methods as function of the global approximation error for the two cases of the analytic magnetic field example in Eq.~(\ref{ex1_mag}).
    The FLOPS are calculated using CountFLOPS code \cite{hangFlops}.
    Both methods exhibit algebraic type scaling for small ${\rm Err}$. However, the increase of the computational complexity is considerably higher for the local splines method that requires several orders of magnitude more operations to achieve comparable small errors. 
}
    \label{flop2_ex1}
\end{figure}

Although the cubic spline and the DivFree-GLS methods can in principle achieve comparable small errors by increasing $N_{\rm train}$ and $M_{\rm term}$ respectively, the complexity costs are quite different. Figure~\ref{flop2_ex1} shows the online FLOPS needed in the two methods for achieving comparable small global approximation errors, ${\rm Err}$. 
For the two magnetic field perturbation cases considered, it is observed that the DivFree-GLS method significantly outperform the cubic spline as ${\rm Err}$ is reduced.  
In these computations, the splines coefficients are pre-computed on the training data set at the nodes of the grid. However, to make a closer mesh-free comparison with the DivFree-GLS method, the search algorithm in the spline evaluation does not assume a structured mesh, which results on a  $\mathcal{O}(N_{\rm test}  \times N_{\rm train} )$ operation. 
Note that the online complexity cost of the cubic spline interpolation method is similar in all cases because the local interpolation method uses the entire training dataset to interpolate the magnetic field regardless of the complexity of the magnetic field. In contrast, the DivFree-GLS approximation method is sensitive to the complexity of the magnetic field and this reduces the FLOPS needed for the unperturbed case by an order of magnitude compared to the perturbed cases.

    The FLOPS are calculated using CountFLOPS code \cite{hangFlops}.
    Both methods exhibit algebraic type scaling for small ${\rm Err}$. However, the increase of the computational complexity is considerably higher for the local splines method that requires several orders of magnitude more operations to achieve comparable small errors. 

\vspace{0.5cm}

To quantify the error in the divergence-free constrain we use
${\rm Div}_{\rm Err}:=(L/B) \nabla \cdot \widehat{\boldsymbol{\mathfrak{B}}}$, where the normalization constants $L$ and $B$ correspond to typical length and magnetic field scales. 
Table \ref{tb_quadrature} compares ${\rm Div}_{\rm Err}$ in the DivFree-GLS and the cubic spline interpolation methods, using $L=a=0.5$ and $B=\mathfrak{B}_0=2$. In all cases the divergence is computed using a central difference method with $\Delta r = \Delta \theta = \Delta \zeta =  10^{-5}$, resulting in a finite-difference error of order $\sim 10^{-10}$ which is significant less than the errors resulting from the interpolation. 
For both methods, the divergence is evaluated at $N_{\rm test} = 10^4$ random spatial locations and the global error is assessed by computing the mean, $\mu$, and standard deviation, $\sigma$, of the set 
$\{ {\rm Div}_{\rm Err}\}$. 
It is observed that the DivFree-GLS method
reaches errors of order $10^{-6}$ with a moderate number of terms in the expansion, whereas the local splines requires fairly large values of $N_{\rm train}$ to achieve similar errors. 

\begin{table}[h!]
\footnotesize
\renewcommand{\arraystretch}{1}
\centering
 \caption{
Mean, $\mu$, and standard deviation, $\sigma$, of  divergence-free constrain error 
${\rm Div}_{\rm Err}:=(L/B) \nabla \cdot \widehat{\boldsymbol{\mathfrak{B}}}$
with $L=a=0.5$ and $B=\mathfrak{B}_0=2$,
sampled  at $10^4$ random spatial locations in  the DivFree-GLS method and the cubic spline method for different values of $M_{\rm term}$ and $N_{\rm train}$ respectively. The DivFree-GLS method achieves mean errors about one order of magnitude smaller than splines, and these errors do not depend strongly on $M_{terms}$.
}
 \label{tb_quadrature}
 \vspace{0.2cm}
  \begin{tabular}{c c c c c c}
  \hline 
  \hline 
  \multicolumn{6}{c}{DivFree-GLS} \\
  \hline
 \multicolumn{3}{c}{one-mode perturbation} & \multicolumn{3}{c}{two-mode perturbation} \\
  \hline
$M_{\rm term}$ &  $\mu$ & $\sigma$ &  $M_{\rm term}$  & $\mu$ & $\sigma$ \\
  \hline
 53 &5.82e-06 & 1.05e-05 & 53 & 2.62e-05 & 2.77e-05\\
  \hline
 72 &5.77e-06 &  1.70e-05 & 108 & 3.35e-05 & 3.71e-05\\
  \hline
131& 5.20e-06 & 1.72e-05 & 181 &  1.81e-05 & 3.23e-05\\
  \hline
170&  1.54e-06 & 2.73e-06 & 222 & 5.60e-06 & 4.10e-06\\
  \hline
197& 1.45e-06 & 2.62e-06 & 263 &  5.35e-06 & 3.93e-06\\
  \hline
  \hline
  \multicolumn{6}{c}{cubic spline} \\
  \hline
 \multicolumn{3}{c}{one-mode perturbation} & \multicolumn{3}{c}{two-mode perturbation} \\
  \hline
$N_{\rm train}$ & $\mu$ & $\sigma $ &  $N_{\rm train}$ & $\mu$ & $\sigma$ \\
  \hline
 $4^3$ &1.15e-03 & 8.78e-04 & $4^3$ & 1.67e-03 & 1.28e-03\\
  \hline
 $8^3$ &1.73e-04 & 2.23e-04 & $8^3$ & 9.66e-04 & 9.83e-04\\
  \hline
 $16^3$ &1.22e-05 & 1.85e-05 & $16^3$ & 5.71e-05 & 9.71e-05\\
  \hline
 $32^3$&6.10e-06 & 5.35e-06 & $32^3$ & 1.22e-05 & 1.03e-05\\
  \hline
 $64^3$&6.02e-06 & 6.00e-06 & $64^3$ & 1.09e-05 & 7.89e-06\\
  \hline
  \hline
  \end{tabular}
\end{table}

\subsubsection{Magnetic field lines orbits}
One of the main uses of the magnetic field line interpolation is in the computation of magnetic field line trajectories defined by $d {\bf r}/ds = {\bm B}({\bf r})$, which in toroidal coordinates reduces to 
\bq\label{odetoro}
\frac{dr}{ds} = B_r, \quad \frac{d\theta}{ds} = \frac{B_\theta}{r}, \quad \frac{d\zeta}{ds} = \frac{B_\zeta}{R_0 + r\cos{\theta}} \, ,
\eq
where $s$ parameterizes the trajectory. The numerical solution of Eq.~(\ref{odetoro}) requires the successive evaluations of the magnetic field at spatial locations along the field line trajectory. In this section we study how the accuracy and efficiency of the interpolation method affects this task which, depending on the numerical method used and the length and number of field lines, can be computational expensive. 
In all the numerical experiments presented here we used an explicit Runge-Kutta method \cite{shampine1997matlab}. 

In the DivFree-GLS method we use $J=15$ and $\{w_1,w_2,w_3\} = \{1,3,1\}$. For the one-mode perturbation case with amplitude $\epsilon_{21} =10^{-4}$, the expansion of the global approximation of $\widehat{\boldsymbol{\mathfrak{B}}}$ has $M_{\rm term} = 197$  ($M_r = 88$,  $M_\theta = 108$, and $M_\zeta = 1$). 
For the two-mode perturbation case with amplitude $\epsilon_{21} =2\times 10^{-4}$,
$M_{\rm term} = 293$ ($M_r = 126$,  $M_\theta = 166$, and $M_\zeta = 1$). Note that because $\delta \boldsymbol{\mathfrak{B}}$ has no $\zeta$ component, $M_\zeta=1$ in all the cases.

\begin{figure}[h!]
\center
    \includegraphics[scale =0.41]{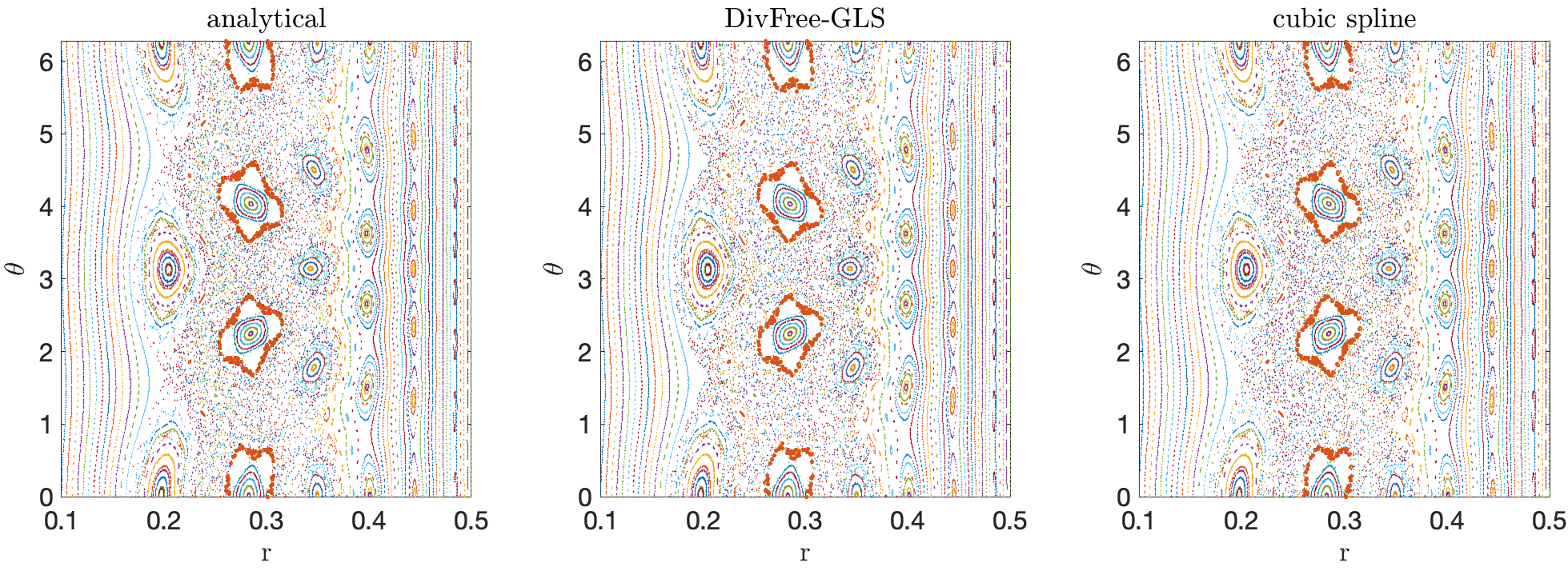}
    \caption{Comparison of Poincar\'e sections 
    for the magnetic field with  a two-mode perturbation with amplitudes  $\epsilon_{21}= \epsilon_{31}= 2 \times 10^{-4}$.
Left panel: interpolation-free direct evaluation using the analytical magnetic field model. Central panel: DivFree-GLS interpolation method. Right panel: local cubic spline interpolation method. All the Poincare plots use the same number of piercings in the $\zeta = 0$-plane and the same initial conditions displayed using the  same colors scheme. 
The results of the three methods are practically indistinguishable, but as shown in Table~\ref{table:1}, the DivFree-GLS method is about two orders of magnitude faster that the cubic splines. 
    }
    \label{poin_mode2_2e04}
\end{figure}

As the Poincare plots in Figure \ref{poin_mode2_2e04} show, there is very good agreement between the DivFree-GLS and the cubic spline methods which are practically indistinguishable from the analytical method that benefits from the interpolation-free exact evaluation of the magnetic field using the model in Eq.~(\ref{ex1_mag}). However, as Table \ref{table:1} shows, there is a significant difference  in the running wall-clock time (measured using the Matlab function ``tic/toc")  among the different methods. As expected, the analytical method that requires the direct evaluation of the simple functions in Eqs.~(\ref{poloidal})-(\ref{h_fcn}) is the fastest and pretty much independent of the type of perturbation. On the other hand, the DivFree-GLS is about two orders of magnitudes faster than the Matlab implementation of the local cubic spline method while tracking the magnetic field orbit in the ODE solver.

\begin{table}[h!]
\footnotesize
\centering
\caption{
Comparison of running times (evaluated using the ``tic/toc" Matlab function) for the computation of the Poincar\'e sections for one-mode and two-mode perturbation cases using the direct evaluation of the analytical field, and the DivFree-GLS and Matlab cubic spline interpolation
methods, for 110 initial conditions. In the cases considered, the DivFree-GLS method is about two orders of magnitude faster than the cubic splines. 
}
\label{table:1}
\vspace{0.3cm}
\begin{tabular}{ |p{3cm}||p{1.5cm}|p{2cm}|p{2cm}|p{3.5cm}|}
  \hline
\multicolumn{5}{|c|}{Poincar\'e sections for one-mode perturbation case} \\
 \hline
Method  &analytical  &DivFree-GLS & cubic spline & spline/DivFree (ratio)\\
 \hline
Running time  (sec)&  2.23 &  185.56 & 1.6987e+04 & 91.55\\
\hline
\multicolumn{5}{|c|}{Poincar\'e sections for two-mode perturbation case} \\
 \hline
Method &analytical & DivFree-GLS & cubic spline &  spline/DivFree (ratio)\\
 \hline
Running time (sec) &  4.55&  350.62 & 2.8920e+04 &82.48\\
\hline
\end{tabular}

\end{table}

\subsection{Numerically generated magnetic field data}

The use of analytical magnetic field models in the previous subsection allowed the efficient generation of large ensembles of training and test data sets to perform 
a  systematic error analysis of the proposed interpolation method. However, in most cases of practical interest, analytical models are not available and the training data sets are usually obtained by running complex magneto-hydrodynamic (MHD) codes. To evaluate the performance of the DivFree-GLS method in this case, we use magnetic data produced by the M3D-C$^1$ \cite{jardin2008m3d} and the NIMROD  \cite{Sovinec04} codes. 
Since both codes provide the magnetic field data in cylindrical coordinates,  we use the DivFree-GLS as described in \ref{sec:cylindrical}. All tests of the DivFree-GLS method use randomly scattered data.

\subsubsection{M3D-C$^1$ magnetic fields}\label{sec:cyl}

The magnetic field in this example was computed using the M3D-C$^1$ code and it models a planned disruption for thermal quench mitigation triggered by the injection of a cyrogenic Ar pellet into an otherwise stable toroidal configuration.
The MHD simulation is initialized with an axisymmetric experimental reconstruction using EFIT \cite{Lao90} of DIII-D discharge $\#160606$ \cite{Lyons20}.
Non-axisymmetic, nonlinear simulations evolve the extended-MHD equations as described in Ref. \cite{Lyons19} and references therein. 
M3D-C$^1$ uses scalar fields representing the magnetic fluxes, allowing $\nabla\cdot{\bm B}=0$ to be exactly satisfied in the calculation.


In this and the following subsection, we compare the DivFree-GLS interpolation method to field evaluation using the FusionIO (FIO) library \cite{Ferraro18} and interpolation using the PSPLINE library \cite{PSPLINE10}. The FIO library was developed to evaluate simulation fields in the native spatial representation of the M3D-C$^1$ simulation. This entails first performing a search to find a given physical location in the logical space of the unstructured mesh. For M3D-C$^1$, the search is assisted by the use of a hint that remembers the previous location in the logical space that is searched first before searching two levels of adjacent elements. If this preliminary search is unsuccessful a full search of the logical space is performed. Once the correct element is found, the evaluation of the reduced-quintic polynomial is straightforward. The PSPLINE library is used to interpolate fields from a uniformly-spaced grid using cubic splines. We use the FIO library to evaluate magnetic fields on this uniformly-spaced grid.
The training set consists of 
$N_{\rm train} = 33 \times 10^4$
elements obtained by sampling the numerically computed non-axisymmetric magnetic field. In the DivFree-GLS method the training data was randomly scattered.
The multi-index sets $\mathcal{J}_R$, $\mathcal{J}_\phi$ and $\mathcal{J}_Z$ are defined using the total degree method in Eq.(\ref{eq_total}) with  $J = 18$, $\{w_1,w_2,w_3\} = \{1,2,1\}$ and $M_{\rm term} = 2608$ terms ($M_R = 882$,  $M_\phi = 857$, $M_Z = 869$).

\begin{figure}[h!]
\center
    \includegraphics[scale =0.41]{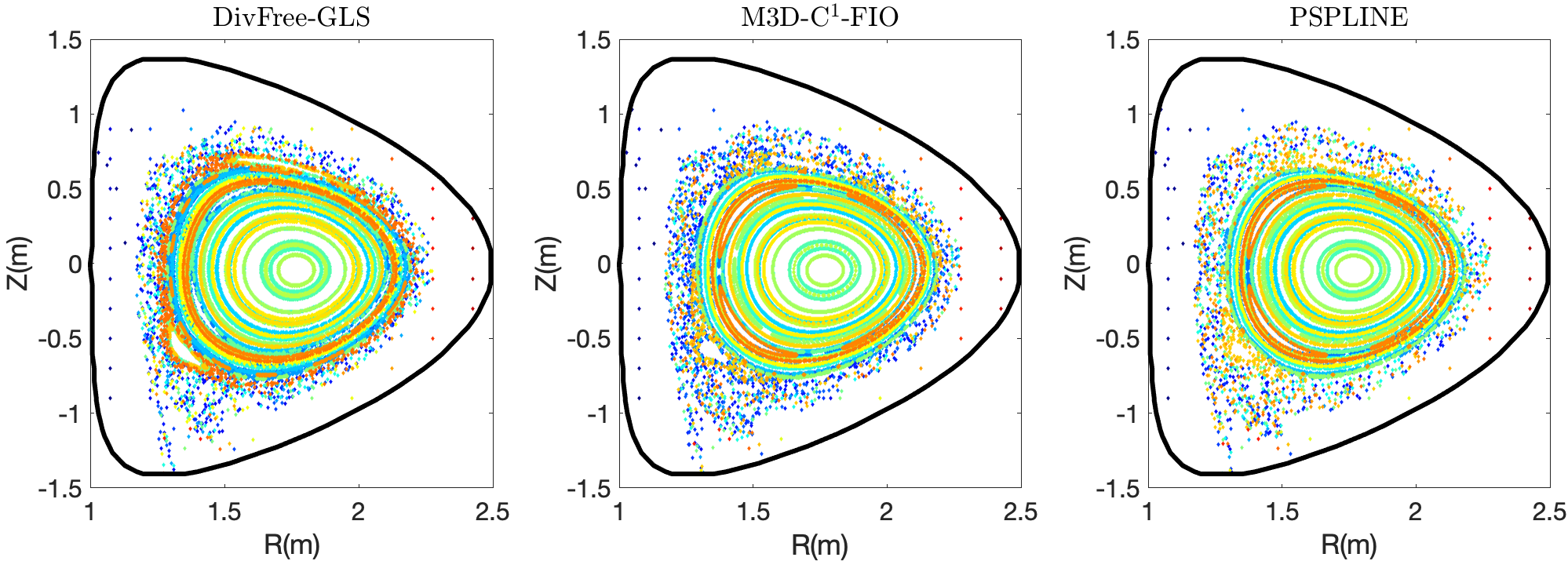}
    \caption{
    Comparison of Poincare sections at $\phi = \frac{\pi}{2}$ computed using the proposed DivFree-GLS method (left panel), the M3D-C$^1$-FIO (center panel), and the PSPLINE library (right panel). The magnetic field was computed using the M3D-C$^1$ code. Shown here is the $t = 1.50$ ms non-axisymmetric magnetic field case. The DivFree-GLS method used randomly scattered training data. Practically identical results (not shown) were obtained for data on an structured grid. The FIO and PSPLINES used structured training data. Except for small differences around the last closed surface, there is very good agreement among the three methods considered. In addition, as shown in Table~\ref{tb_m3dc1}, the DivFree-GLS method keeps a small divergence error. 
    }
    \label{t60ls}
\end{figure}

Figure~\ref{t60ls} presents the comparison of Poincar\'e sections of the magnetic field lines at an intermediate time of the computation. 
Very good agreement is observed among the DivFree-GLS, the M3D-C$^1$-FIO, and PSPLINE interpolation methods, even in the regions where the magnetic field exhibits islands and stochasticity. 
Note that the M3D-C$^1$ data is given inside a ``D-shape" domain which corresponds to the geometry of the cross section of the DIII-D tokamak. In principle this can give rise to inaccuracies near the boundaries when using local interpolation methods that require the use of ``ghost data" outside the boundary usually padded with zero values. This is not the case in the DivFree-GLS which handles  arbitrary training sets $\mathcal{S}$ globally with no need to impose a special local treatment at the boundary. In addition, as reported in Table~\ref{tb_m3dc1}, the DivFree-GLS method has a small divergence-free constrain error.



\begin{table}[h!]
\footnotesize
\renewcommand{\arraystretch}{1}
\centering
 \caption{
 Mean, $\mu$, and standard deviation, $\sigma$, of  divergence-free constrain error, 
${\rm Div}_{\rm Err}:=(L/B) \nabla \cdot \widehat{\boldsymbol{\mathfrak{B}}}$,
with $L=0.67$ and $B=2$. 
Error sampled  at $10^4$ random spatial locations of a M3D-${\rm C}^1$ simulation  for different values of $M_{\rm term}$ using  the DivFree-GLS method with randomly scattered training data.
%
}
\vspace{0.2cm}
\label{tb_m3dc1}
  \begin{tabular}{c c c c c c}
  \hline 
  \hline 
 \multicolumn{6}{c}{m3dc1 magnetic field} \\
  \hline
$M_{\rm term}$ &375 &1092 & 1444&  1917& 2608 \\
\hline
$\mu$  &4.31e-03& 4.49e-04 &2.27e-04 &2.59e-04 &1.74e-04 \\
\hline
$\sigma$ & 2.32e-03 & 4.69e-03 & 1.22e-03 & 3.13e-04 & 3.47e-04\\
  \hline
  \hline
  \end{tabular}
\end{table}

\subsection{NIMROD magnetic fields}

The NIMROD magnetic fields explored in this section are initialized according to the methodology in Ref.~\cite{cornille2022computational} from runaway electron deconfinement experiments in the Madison Symmetric Torus (MST) reported in Ref.~\cite{Munaretto20}. 
The NIMROD spatial representation uses structured, C$^0$-continuous, Lagrange spectral elements in the poloidal plane and a finite series of Fourier modes in the toroidal direction. The simulation of interest in this section uses a $32\times32$ poloidal mesh of 2D quadrilateral, bicubic elements and $0\ge n\ge10$ toroidal Fourier modes. 
Perturbation fields are included that model an externally-applied 3D field with a well-defined poloidal mode number $m=3$ and spectrum of toroidal mode number characteristic of the poloidal gap on MST, with form defined by Eq.~(2) in Ref.~\cite{cornille2022computational}. 
Because NIMROD solves for the magnetic field in various configurations, the numerical formulation does not identically have a divergenceless magnetic field as in M3D-C$^1$, but uses a “divergence cleaner". 

The training datasets  $\boldsymbol{\mathcal{D}}_R^{\rm train}$, $\boldsymbol{\mathcal{D}}_\phi^{\rm train}$ and $\boldsymbol{\mathcal{D}}_Z^{\rm train}$, are the magnetic field values at  $\{(R_n,\phi_n,Z_n)$, $n=1, \dots N_{\rm train}\}$ with 
$N_{\rm train}=51 \times 10^2$ elements.
In the DivFree-GLS method the training data was randomly scattered.
As in the M3D-C$^1$ case, we use the total-degree method to determine the multi-index sets $\mathcal{J}_R$, $\mathcal{J}_\phi$ and $\mathcal{J}_Z$ with  $J = 18$ and $\{w_1,w_2,w_3\} = \{1,2,1\}$ resulting in an  expansion of the approximation $\widehat{\boldsymbol{\mathfrak{B}}}$ containing $M_{\rm term} = 2782$ terms ($M_R = 932$,  $M_\phi = 935$, $M_Z = 915$).

\begin{figure}[h!]
\center
    \includegraphics[scale =0.41]{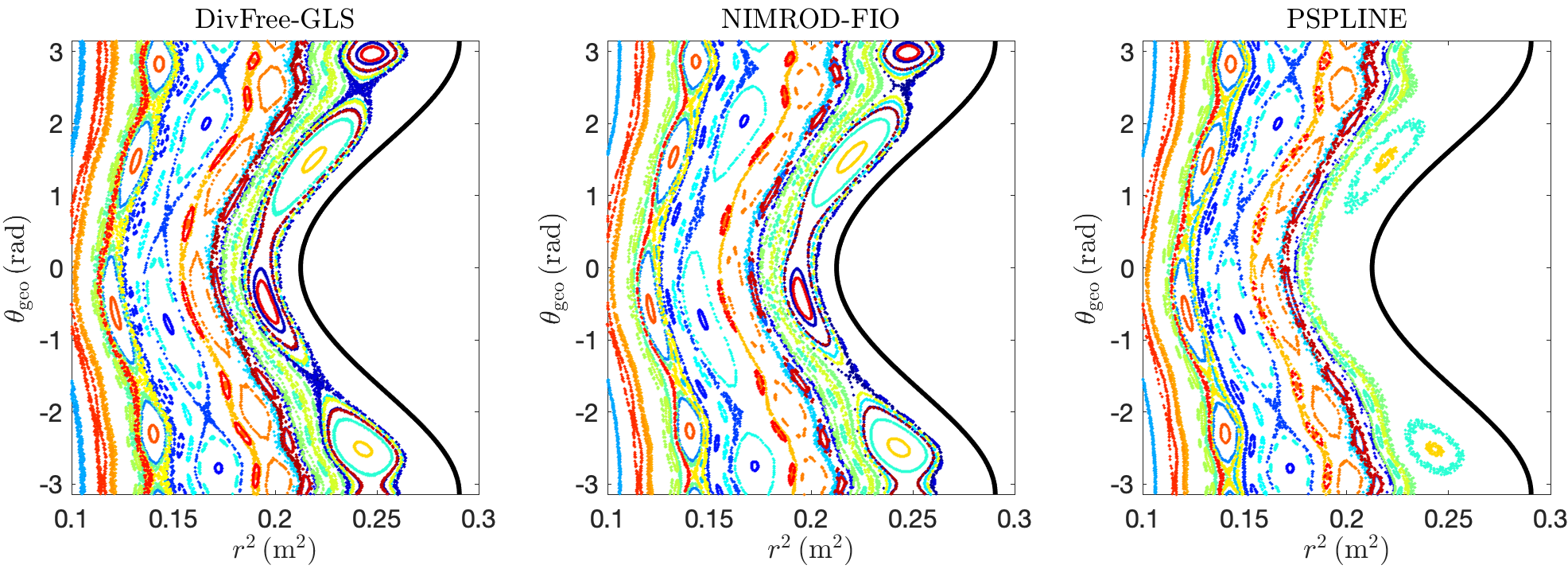}
    \caption{
    Comparison of Poincare sections computed using the proposed DivFree-GLS method (left panel), the NIMROD-FIO (center panel), and the PSPLINE library (right panel). 
    The magnetic field was computed using the NIMROD code and to ease the comparison flux-type coordinates are used with the square of the minor radius in the x-axis and the poloidal angle in the y-axis. 
    The DivFree-GLS method used randomly scattered training data. Practically identical results (not shown) were obtained for data on an structured grid. The FIO and PSPLINES used structured training data.
    Very good agreement is found between the DivFree-GLS method and the NIMROD-FIO. The discrepancies found in the PSPLINE near the boundary (dark solid line) can be eliminated by doubling the size of the training set at the expense of increasing the computational complexity. 
    }
    \label{POin_nim}
\end{figure}

Figure.~\ref{POin_nim} compares the Poincar\'e sections computed using the 
DivFee-GLS method, the NIMROD-FIO, and the PSPLINES library. For theses calculations the FIO library has been extended to evaluate fields in the spatial representation of NIMROD simulations. If the search fails to find the logical location using the hint of the previous location, a Newton-Raphson method is used to find the logical location given the physical location. Similar to the previous case with M3D-C$^1$, once the logical location is found the evaluation of the polynomial is straightforward. To ease the comparison, in this case we use flux-type coordinates with the x-axis corresponding to the square of the minor radius and the y-axis corresponding to the poloidal angle. 
Very good agreement is found between the DivFree-GLS and NIMROD-FIO methods. However, the PSPLINE library exhibits some discrepancies in the vicinity of the boundary. Further calculations (not shown) indicate that these boundary errors can be reduced by increasing the size of the training data set. In particular, increasing $N_{\rm train}$ from $N_{\rm train}=100 \times 51 \times 100$ to 
$100 \times 101 \times 100$ yield PSPLINE results in agreement to the DivFree-GLS and NIMROD-FIO methods. Although this is reassuring, doubling  the training data set doubles the computational complexity of the spline computation. 
Like in the previous example, as shown in Table~\ref{tb_nimrod} the  DivFree-GLS method preserves the divergence-free condition
well. 



\begin{table}[h!]
\footnotesize
\renewcommand{\arraystretch}{1}
\centering
 \caption{
Mean, $\mu$, and standard deviation, $\sigma$, of  divergence-free constrain error, 
${\rm Div}_{\rm Err}:=(L/B) \nabla \cdot \widehat{\boldsymbol{\mathfrak{B}}}$,
with $L=0.5$ and $B=0.14$. 
Error sampled  at $10^4$ random spatial locations of a NIMROD simulation  for different values of $M_{\rm term}$ using  the DivFree-GLS method with randomly scattered training data.
}
\label{tb_nimrod}
\vspace{0.2cm}
  \begin{tabular}{c c c c c c}
  \hline 
  \hline 
 \multicolumn{6}{c}{nimrod magnetic field} \\
  \hline
$M_{\rm term}$ &390 &875& 1629 & 2178& 2782 \\
\hline
$\mu$  &1.33e-02& 2.31e-03 &2.08e-03 &1.93e-03 &1.97e-03 \\
\hline
$\sigma$ & 7.25e-03 & 4.98e-03 & 3.45e-03 & 2.26e-03 & 2.07e-03\\
  \hline
  \hline
  \end{tabular}
\end{table}

\begin{figure}[h!]
\center
    \includegraphics[scale =0.41]{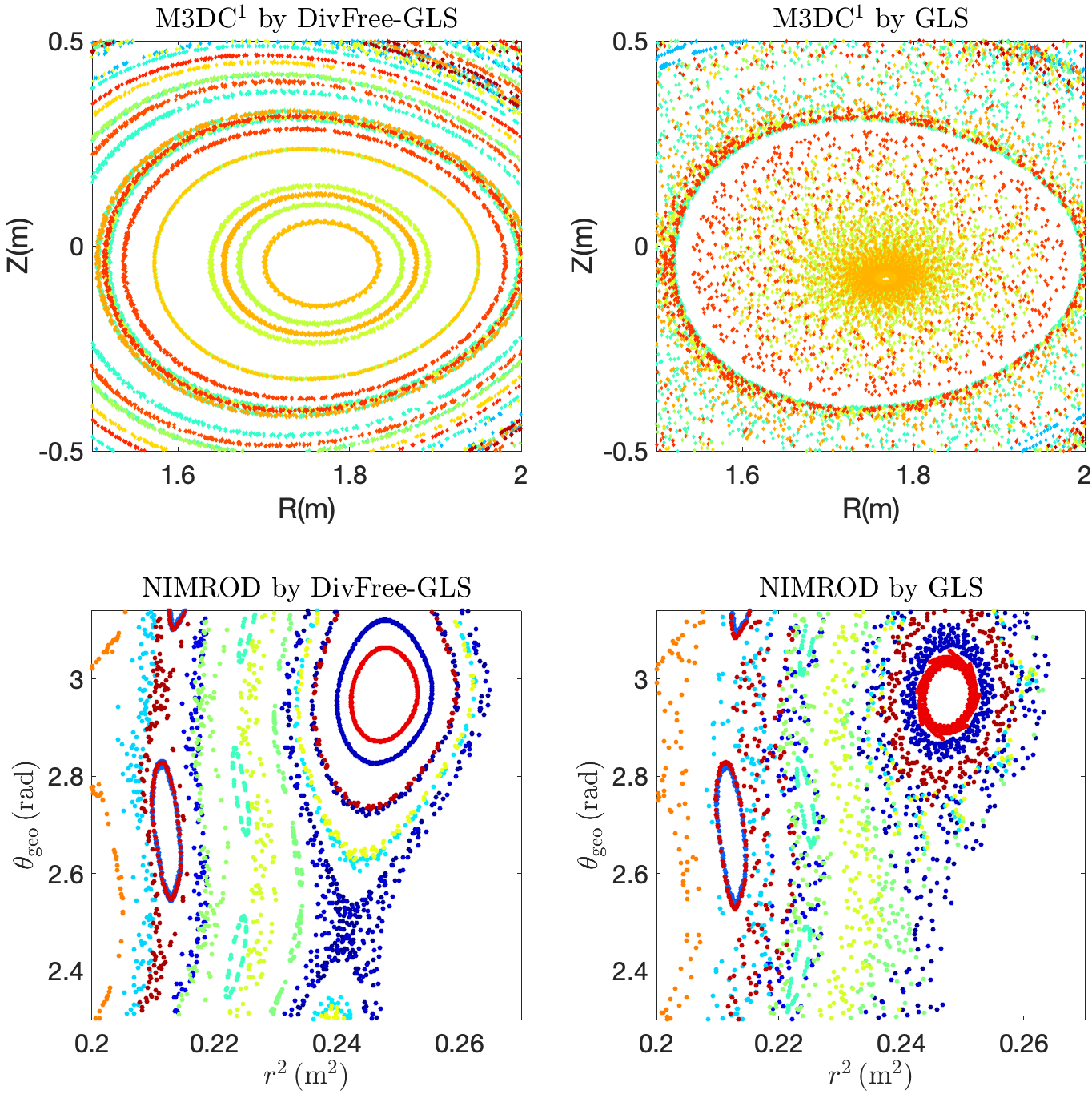}
    \caption{
    Poincare sections illustrating the critical role of the divergence-free constrain. The left column shows the Poincare section computed using the proposed DivFree-GLS and the right column shows the results of a global least-square minimization without the divergence-free constrain. The top (bottom) row corresponds to the M3DC$^1$ (NIMROD) case. Not enforcing the divergence free condition leads to unphysical results including the spiraling of integrable magnetic field trajectories towards the center, the blurring of small islands and the artificial increase of magnetic stochasticity. 
    }
    \label{div_nondiv}
\end{figure}

Figure~\ref{div_nondiv} illustrates 
the importance of the divergence-free constrain in the computation of Poincare plots. In particular, interpolations using global least squares without the divergence-free constrain in Eq.~(\ref{div_mat})  can lead to unphysical results for both integrable (top left panel) and stochastic (bottom left panel) orbits.

\section{Summary and conclusions}\label{sec:clu}

We have presented a novel magnetic field interpolation method. The starting point is a training data set obtained by sampling a ground truth magnetic field, $\boldsymbol{\mathfrak{B}}$, on a set of $N_{\rm train}$ points in a  spatial domain, ${\mathcal S}$. The spatial location of the sample data does not necessarily need to correspond to the nodes of a coordinate grid and  ${\cal S}$ can have an arbitrary shape and dimension. The fundamental idea of the method does not depend on the coordinate system, but the actual implementation does. Here we discussed implementations in toroidal (main text) and cylindrical (appendix) coordinates. The method is based on the construction of an approximate magnetic field, $\widehat{\boldsymbol{\mathfrak{B}}}$, consisting of a linear superposition of orthogonal functions. Once this is done, the interpolation reduces to the direct evaluation of the function $\widehat{\boldsymbol{\mathfrak{B}}}$.
The expansion coefficients, $\alpha^k_{\bf j}$, are obtained by solving, in the least-squares sense, an over-determined linear system of equations. In particular, the values of
$\alpha^k_{\bf j}$, minimize an error cost function, ${\cal E}$, defined as the $L^2$ norm of the difference between the ground truth and the approximate magnetic field, as well as the divergence-free condition, on the training data set. The method is global in the sense that  uses all the data at once to minimize the overall error and the expansion functions are well-defined in the whole space.
 The divergence-free condition is incorporated as a ``soft" constrain in the cost function. 
 Therefore, although the method can achieve arbitrary small errors in the magnetic field divergence, the divergence condition is not preserved exactly to machine precision.  
 
The main parameter controlling the accuracy of the proposed DivFree-GLS method is the number of terms, $M_{\rm term} \ll N_{\rm train}$, used in the expansion, and numerical simulations show an exponential decay of the global approximation error, ${\rm Err}$, with $M_{\rm term}$. As a way of comparison, for the same data, the local spline method exhibits a less favorable algebraic decay of ${\rm Err}$ with $N_{\rm train}$. 
As the error decays, the computational cost of the local splines method increases at a significantly higher rate compared to the  DivFree-GLS method. This reflects in the computational cost of Poincare sections for which the DivFree-GLS method exhibits about two orders of magnitude faster performance without compromising accuracy. 
Using the analytical properties of the basis functions, the DivFree-GLS method also provides an accurate and efficient interpolation of the  derivatives of the magnetic field, needed, for example, in the integration of guiding center orbits. 

Numerical tests with magnetic field data computed using the state-of-the-art MHD codes M3DC$^1$ and NIMROD in general show very good agreement between the DivFree-GLS method and the standard Fusion-IO and PSPLINES libraries even in the presence of magnetic filed stochasticity. However, in the NIMROD case, PSPLINES exhibits some discrepancies at the boundaries that can be resolved by increasing the size of the training data set at the expense of increasing the computational cost. For both M3DC$^1$ and NIMROD, the mean error on the divergence in the DivFree-GLS method is of the order of $\sim 10^{-4}$ and $\sim 10^{-3}$ respectively. Most importantly, this error tends to stay small, even when the number of terms, $M_{\rm term}$, in the  expansion is reduced by an order of magnitude. 
It is interesting to observe that the proposed method could also be viewed as a  ``divergence-cleaning" method in the sense that given a discrete data set obtained by sampling a vector field, ${\bf B}$, such that $|\nabla \cdot {\bf B}| = \delta>0$,  it
reconstructs a field, $\widehat{\bf B}$, that approximates the given data optimally with arbitrarily small divergence,
$| \nabla \cdot \widehat{\bf B}| = \epsilon \ll \delta$.
Note also that the divergence error can be targeted and independently improved by introducing a scaling parameter $\lambda>1$ into the divergence-free constrain in the last row, $[ \lambda {\bf A}_r'\,\,\, \lambda {\bf A}_\theta'\,\,\, \lambda {\bf A}_\zeta']$, of the cost function matrix in  Eq.~(\ref{mat_con1_2}). 

In principle, our method can also be used to interpolate magnetic fields including other constrains in addition to the divergence-free condition. A case of particular interest is the curl-free constrain, $\nabla \times {\bf B}=0$, of interest to the interpolation of current-free, static magnetic fields, or irrotational, incompressible fluids. However, in practice this might be computationally expensive because in the proposed method these constrains couple the three independent (one for each magnetic field component) least-squares problems leading to large computational memory demands, specially for large training data sets. 

An advantage of the DivFree-GLS method over more traditional interpolation methods is that no search algorithm is needed, which allows vectorization to be used in all cases. The FIO method, as used with M3D-C1 and NIMROD fields, must search through unstructured and irregular meshes requiring the inclusion of conditional statements that preclude the use of vectorization. Furthermore, search algorithms using the PSPLINE library are limited to regular meshes, otherwise, conditional statements are required. The direct field evaluations using the DivFree-GLS method contain no conditional statements and can take advantage of vector operations.


\section{Acknowledgements}

We thank B.~Lyons and S.~Jardin for proving the data of the  M3D-C$^1$ simulations, N.~Ferraro and M.~Cianciosa for help with the Fusion-IO interpolation, and B.~Cornille and C.~Sovinec for proving the data of the NIMROD calculations and help with the interpolation libraries. 
This material is based upon work supported by the U.S. Department of Energy, Office of Science, Office of Advanced Scientific Computing Research and Office of Fusion Energy Science, Scientific Discovery through Advanced Computing (SciDAC) program, at the Oak Ridge National Laboratory, which is operated by UT-Battelle, LLC, for the U.S. Department of Energy under Contract DE-AC05-00OR22725.


\section*{References}
\bibliography{B_ref}
\bibliographystyle{abbrv}

\appendix
\section{The DivFree-GLS algorithm in cylindrical coordinates}\label{sec:cylindrical}
In cylindrical coordinates the magnetic field ${\bm B}$ has the form 
\bq\label{mag_cylin}
        {\bm B}(R,\phi,Z)=B_R(R,\phi,Z) {\bf e}_R+ B_\phi(R,\phi,Z) {\bf e}_\phi + B_Z(R,\phi,Z) {\bf e}_Z \,,
\eq
where
$x=R \cos \phi$, $y=R \sin \phi$, and $z=Z$, and  ${\bm e}_R$, ${\bm e}_\phi$, ${\bm e}_Z$ are the corresponding unit basis vectors. The domains of the variables are $R \in [R_{min},R_{max}]$, $\phi \in [0, 2 \pi)$ and  $Z=[Z_{min}, Z_{max}]$, and the training dataset is
\bq\label{givencylin}
\begin{aligned}
\boldsymbol{\mathcal{D}}_k^{\rm train} := \left\{(R_n, \phi_n, Z_n, {B}_k(R_n,\phi_n,Z_n))\;|\; (R_n, \phi_n, Z_n) \in \mathcal{S} \text{ for } n = 1, \ldots, N_{\rm train}\right\},
\end{aligned}
\eq
where the sub-index $k$ labels the radial, $k = R$, angular, $k=\phi$, and vertical, $k = Z$ components. Since the divergence condition,
\begin{equation}\label{free_cylin}
\begin{aligned}
        \nabla \cdot {\bm B} = & \frac{1}{R}\frac{\partial (RB_{R})}{\partial R}+\frac{1}{R}\frac{\partial B_{\phi}}{\partial \phi} + \frac{\partial B_{Z}}{\partial Z} = 0 \, ,
\end{aligned}
\end{equation}
is relatively simple, 
there is no need to rescale the magnetic field.  

Given the training data set, the goal is to construct 
\begin{equation}
\widehat{\bm B}(R, \phi, Z) := \widehat{B}_R(R,\phi,Z) {\bm e}_R+ \widehat{B}_\phi(R,\phi,Z) {\bm e}_\phi +  \widehat{B}_Z(R,\phi,Z) {\bm e}_Z,
\end{equation}
where $\widehat{B}_R$, $\widehat{B}_\phi$, $\widehat{B}_Z$ are approximations of ${B}_R$, ${B}_\phi$, ${B}_Z$ respectively, and defined by the expansions
\begin{equation}\label{eq_toro_approx}
\begin{aligned}
\widehat{B}_R (R,\phi,Z)& := \sum_{\mathbf{j}\,\in\, \mathcal{J}_R} \alpha^{R}_{\mathbf{j}}\; \psi_{\mathbf{j}}(R,\phi,Z), \\
\widehat{B}_\phi (R,\phi,Z)& := \sum_{\mathbf{j}\,\in\, \mathcal{J}_\phi} \alpha^{\phi}_{\mathbf{j}}\; \psi_{\mathbf{j}}(R,\phi,Z),\\
\widehat{B}_Z (R,\phi,Z)& := \sum_{\mathbf{j}\,\in\, \mathcal{J}_Z} \alpha^{Z}_{\mathbf{j}}\; \psi_{\mathbf{j}}(R,\phi,Z), 
\end{aligned}
\end{equation}
where $\mathbf{j} := (j_1, j_2, j_3)$ is a multi-index in the index sets $\mathcal{J}_R$, $\mathcal{J}_\phi$ or $\mathcal{J}_Z$, $\psi_{\mathbf{j}}(R,\phi,Z)$ are global basis functions associated with the index $\mathbf{j}$, and $\alpha_{\mathbf{j}}^R$, $\alpha_{\mathbf{j}}^\phi$, $\alpha_{\mathbf{j}}^Z$ are the corresponding coefficients. In cylindrical coordinates, the global basis functions are given by \[
\psi_{\mathbf{j}}(R,\phi,Z) := P_{j_1}(R)F_{j_2}(\phi)P_{j_3}(Z),
\]
where $P_j$ is the $j$th order Legendre polynomial and $F_j$ is the $j$th order trigonometric function.
The multi-index sets $\mathcal{J}_R$, $\mathcal{J}_\phi$ and $\mathcal{J}_Z$ are defined using the  full tensor product
\begin{equation}\label{tensor_cylin}
\mathcal{J}^{\rm full-tensor}:= \{\mathbf{j} = (j_1,j_2,j_3) \,\,|\,\,  0\leq j_1\leq J_1, |j_2|\leq J_2, j_3\leq J_3\},
\end{equation}
where $J_1$, $J_2$, $J_3$ are given parameters,  or the total degree method
\begin{equation}\label{total_cylin}
\mathcal{J}^{\rm total-degree}:= \{\mathbf{j} = (j_1,j_2,j_3) \,\,|\,\,  w_1 j_1 + w_2 |j_2| + w_3 j_3 \leq J\},
\end{equation}
where the weights $w_1$, $w_2$, $w_3$ and $J$ are given parameters. 
The number of elements in $\mathcal{J}_R$, $\mathcal{J}_\phi$ and $\mathcal{J}_Z$ are denoted by 
$M_{R}$, $M_{\phi}$ and $M_{Z}$ respectively. 

As before, the expansion coefficients are determined by imposing the equality of the ground-truth field and the approximation at the training set. For the radial component this gives
\begin{equation}\label{algo_31}
\begin{aligned}
\widehat{B}_R(R_n,\phi_n,Z_n)=\sum_{\mathbf{j}\,\in\, \mathcal{J}_R} \alpha^{R}_{\mathbf{j}}\; \psi_{\mathbf{j}}(R_n, \phi_n, Z_n) = B_R (R_n,\phi_n,Z_n)\, ,
\end{aligned}
\end{equation}
for $n = 1, \ldots, N_{\rm train}$, which can be written in matrix form as
\begin{equation}
\label{mat_form2}
{\bf A}_R{\bm \alpha}_{R} = {\bm b}_R \, .
\end{equation} 
%
where ${\bf A}_R \in \mathbb{C}^{N_{\rm train}\times M_{R}}$ is determined by the global basis functions $\psi_{\mathbf{j}}(R_n,\phi_n,Z_n)$, for $\mathbf{j}\,\in\, \mathcal{J}_R$, $n = 1, \ldots, N_{\rm train}$, the unknown coefficient vector, ${\bm \alpha}_R \in \mathbb{C}^{M_{R}}$, consists of $\alpha^{R}_{\mathbf{j}}$,  $\mathbf{j} \in \mathcal{J}_R$, and ${\bm b}_R \in \mathbb{R}^{N_{\rm train}}$ is the data vector. Similar arguments lead to the conditions for the $\phi$ and $Z$ components
\begin{equation}\label{algo_34}
\begin{aligned}
{\bf A}_\phi {\bm \alpha}_\phi = {\bm b}_\phi\, , \qquad
{\bf A}_Z {\bm \alpha}_{Z} = {\bm b}_Z.
\end{aligned}
\end{equation}
%

%
Finally, like in the toroidal coordinates case, we write the divergence-free condition in Eq.(\ref{free_cylin}) for the approximate magnetic field $\widehat{\bm B}$ in the matrix form, 
\begin{equation}\label{div_mat2}
 \begin{aligned}
 {\bf A}'_R  {\bm \alpha}_{R} + {\bf A}'_\phi {\bm \alpha}_\phi + {\bf A}'_Z {\bm \alpha}_Z = 0 \, ,
\end{aligned}
\end{equation}  
where the matrices ${\bf A}'_R$, ${\bf A}'_\phi$ and ${\bf A}'_Z$ are constructed from $\widehat{\bm B}$ using the properties of the derivatives of the Legendre polynomials and the trigonometric functions. 

Like in the toroidal coordinates case, the linear system Eqs.~\eqref{mat_form2}, \eqref{algo_34}, and \eqref{div_mat2} is in general over-determined and thus we seek a solution in the least squares sense. That is, we define
\begin{equation}\label{mat_con2}
{\cal E} :=\left\|
\begin{pmatrix}
{\bf A}_R & {\bf 0} & {\bf 0}\\
{\bf 0} & {\bf A}_\phi & {\bf 0}\\
{\bf 0}& {\bf 0} & {\bf A}_Z \\
{\bf A}'_R & {\bf A}'_\phi & {\bf A}'_Z 
\end{pmatrix} 
\begin{pmatrix}
{\bm \alpha}_R \\
{\bm \alpha}_{\phi} \\
{\bm \alpha}_{Z}
\end{pmatrix} -
\begin{pmatrix}
{\bm b}_R \\
{\bm b}_\phi \\
{\bm b}_Z \\
{\bf 0}
\end{pmatrix}\right \| \, ,
\end{equation}
and find the ${\bm \alpha}_R$, ${\bm \alpha}_\phi$, and ${\bm \alpha}_Z$ that minimize ${\cal E}$.
\end{document}